\documentclass[fleqn,usenatbib,useAMS]{mnras}

\usepackage{graphicx}	
\usepackage{amsmath}	
\usepackage{amssymb}	
\usepackage{multicol}        
\usepackage{bm}		
\usepackage{pdflscape}	
\usepackage{xcolor}
\graphicspath{{figs/}}

\newcommand{\comp}{c/\omega_{pe}}
\newcommand{\ompi}{\omega_{pe}^{-1}} 
\newcommand{\sigp}{\sigma^{\prime}} 

\usepackage[T1]{fontenc}
\usepackage{ae,aecompl}

\usepackage{newtxtext,newtxmath}

\title[MHD Shock Departure]{Kinetic Simulations of Strongly-Magnetized Parallel Shocks: Deviations from MHD Jump Conditions}

\author[C. C. Haggerty et al.]{
Colby C. Haggerty,$^{1,2}$\thanks{E-mail: colbyh@hawaii.edu (CCH)}
Antoine Bret,$^{3,4}$
Damiano Caprioli$^{2,5}$
\\
$^{1}$Institute for Astronomy, University of Hawai`i, Honolulu, HI 96822, USA\\
$^{2}$Department of Astronomy \& Astrophysics, University of Chicago, Chicago, IL 60637, USA\\
$^{3}$ETSI Industriales, Universidad de Castilla-La Mancha, 13071 Ciudad Real, Spain\\
$^{4}$Instituto de Investigaciones Energ\'{e}ticas y Aplicaciones Industriales, Campus Universitario de Ciudad Real, 13071 Ciudad Real, Spain\\
$^5$Enrico Fermi Institute, The University of Chicago, 5640 S Ellis Ave, Chicago, IL 60637, USA
}

\date{\today}

\pubyear{2021}

\begin{document}
\label{firstpage}
\pagerange{\pageref{firstpage}--\pageref{lastpage}}
\maketitle

\begin{abstract}
Shocks waves are a ubiquitous feature of many astrophysical plasma systems, and an important process for energy dissipation and transfer.
The physics of these shock waves are frequently treated/modeled as a collisional, fluid MHD discontinuity, despite the fact that many shocks occur in the collisionless regime.
In light of this, using fully kinetic, 3D simulations of non-relativistic, parallel propagating collisionless shocks comprised of electron-positron plasma, we detail the deviation of collisionless shocks form MHD predictions for varying magnetization/Alfv\'enic Mach numbers, with particular focus on systems with Alf\'enic Mach numbers much smaller than sonic Mach numbers.
We show that the shock compression ratio decreases for sufficiently large upstream magnetic fields, in agreement with the predictions of \cite{BretJPP2018}.
Additionally, we examine the role of magnetic field strength on the shock front width.
This work reinforces a growing body of work that suggest that modeling many astrophysical systems with only a fluid plasma description omits potentially important physics.
\end{abstract}


\section{Introduction}
In a collisional fluid, dissipation at the front of a shockwave is provided by binary collisions. As a consequence, the shock front is a few mean free path thick \citep{Zeldovich}. 
Yet, in plasma, shock waves can be sustained by collective effects, with a front smaller than the mean free path by several orders of magnitude  \citep{Petschek1958,Buneman1964,Sagdeev66}.

A good example is the bow shock formed at the boundary of the Earth's magnetosphere and the solar wind, in which the shock front thickness is on the order of  100 kilometers, while the proton mean free path is comparable to an AU \citep{PRLBow1,PRLBow2}.

Such shocks are omnipresent in astrophysical systems and are good potential candidates to explain the origin of cosmic rays, gamma-ray bursts and fast radio bursts; as such, they have been under intense scrutiny for decades \citep[e.g.,][]{Blandford1987,Piran2005,Sironi+21}. 
Because of their collisionless characteristics the fluid formalism cannot be straightforwardly applied to study them. 
Consequently, Particle-In-Cell (PIC) simulations, based on a model that evolves the equations of collisionless, kinetic plasmas, has become an important tool employed in the investigation of shocks in recent years 
\citep[e.g.,][]{Hoshino+92,Dieckmann2000,Spitkovsky2008,sironi_spitkovsky_11a,Caprioli2014,Park2016JPhC}.
The fluid properties of these kinetic shock simulations, such as the compression ratio and downstream heating, are often measured by their corresponding MHD predictions, even though the relevance of MHD for collisionless shocks is not straightforward \citep{BretApJ2020,haggerty+2020}. 

In this respect, parallel shocks are particular interesting when comparing the kinetic and MHD picture. In the MHD picture, the strength of a parallel magnetic field does not affect the hydrodynamics of shock; in this configuration, the field and the fluid are completely disconnected (\cite{Lichnerowicz1976}, or \cite{Kulsrud2005}, p. 141). Therefore, any detectable effect of a parallel field on a collisionless shock represents a departure from MHD and must be due to kinetic effects.

In a series of recent theoretical works collisionless (electron-positron) plasma shock hydrodynamics were reconsidered from a kinetic framework \citep{BretJPP2017,BretJPP2018,BretLPB2020}, in which the density jump of a collisionless shocks was predicted deviate from the standard MHD, Rankine Hugoniot jump conditions; namely that an increasing the magnetic field strength in a parallel shock should decrease the compression ratio. In this paper, we constrain this prediction with state-of-the-art PIC simulations, and we examine the effects of increased field strength on the thermal and nonthermal properties of the shock.
The paper is outlined as follows: 
In Sec.~\ref{sec:theo} we review the theory and predictions for the departure from MHD,
in Sec.~\ref{sec:PIC} we discuss the PIC model and the setup for the shock simulations,
in Sec.~\ref{sec:hydro} the fluid properties of the are examined, namely the density, the anisotropic temperature and the entropy, and are shown to be in good agreement with theory, in Sec.~\ref{sec:width}
we detail how the magnetic field strength modified the shock front's width 
, and finally we conclude in Sec.~\ref{sec:conc}.

\begin{figure}
 \includegraphics[width=\columnwidth]{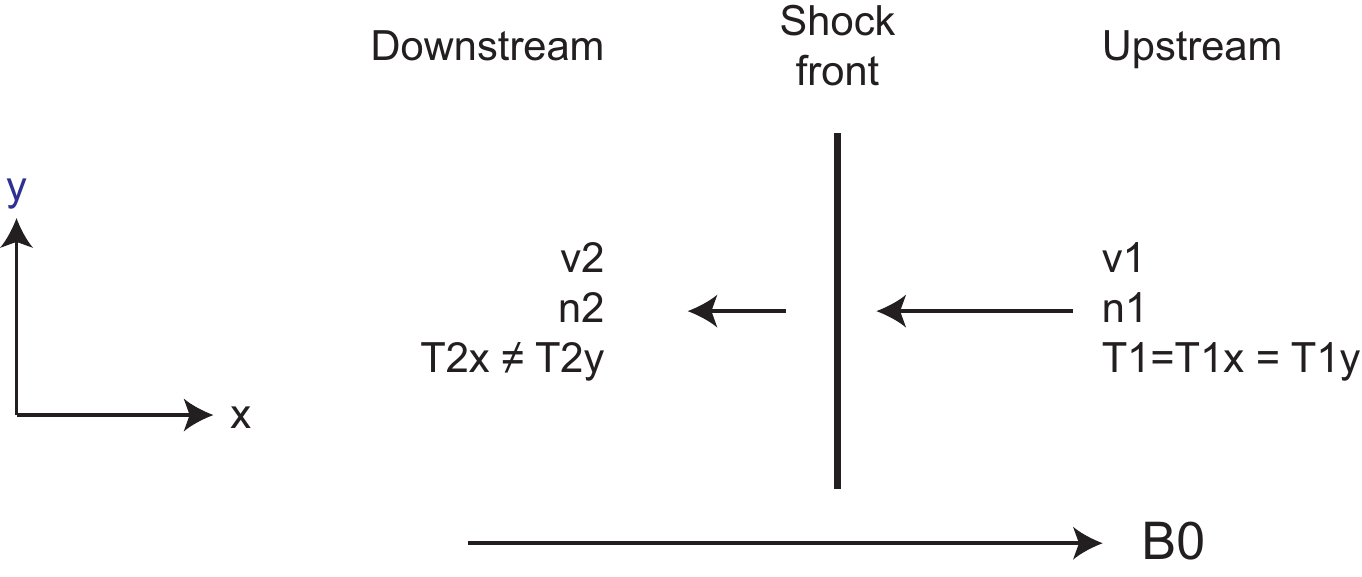}
 \caption{A diagram of the shock system considered, and setup in the simulations, with subscripts 1 and 2 denoting upstream and downstream quantities respectively.}
 \label{fig:setup}
\end{figure}

\section{Theory}\label{sec:theo}
For completeness we briefly summarize the model developed for the density jump in \cite{BretJPP2018}, hereafter BN18.  We considered a non-relativistic shock in an electron-positron (pair) plasma. The upstream is assumed isotropic. The external magnetic field $\bmath{B}_0$ is parallel to the shock normal, as shown in Fig.~\ref{fig:setup}. The upstream density, pressure and temperature are $n_1, P_1$  and $T_1$ respectively. Downstream quantities are labeled with the subscript ``2'', and the downstream are assumed to be anisotropic with $T_{2x}\neq T_{2y}$. We define the downstream anisotropy parameter $A_2$ as,
\begin{equation}\label{eq:A2}
A_2 = \frac{T_{2y}}{T_{2x}}.
\end{equation}
The field strength is measured by the $\sigma$ parameter (magnetization),
\begin{equation}
    \sigma \equiv \frac{B_0^2/8\pi}{\frac{1}{2} n_1 m v_1^2} = \frac{v_{A0}^2}{v_1^2},
    \label{eq:sigma}
\end{equation}
where $v_{A0}$ is the Alfv\'en speed and $v_1$ the upstream velocity in the shock frame.

While the standard MHD jump conditions assume isotropy, several authors have adapted them to account for a downstream pressure anisotropy $A_2$ \citep{Karimabadi95,Vogl2001,Erkaev2000,Gerbig2011}. Yet, in these studies, $A_2$ is a free parameter leaving the problem under-determined and an additional assumption is required to pinpoint $A_2$ in terms of the upstream quantities \citep{Gary1993,BalePRL2009,MarucaPRL2011}.

The additional assumption made in BN18 is that the transit through the shock front is adiabatic in the perpendicular direction while the entropy increase goes into the parallel temperature. Regarding the perpendicular temperatures, the relations derived in \citep{CGL1956} imply their conservation. Together with the macroscopic conservation equations, this assumption allows to fully determine the characteristics of the downstream, including its temperature anisotropy $A_2$.

Depending on the value of $A_2$ thus obtained, the downstream can be firehose unstable if the temperature anisotropy is large relative to the magnetic field strength. For weak fields, i.e., where $\sigma$ smaller than some critical value $\sigma_c$, the downstream is firehose unstable, while for strong enough fields, with $\sigma > \sigma_c$, the downstream remains stable and constitutes the end state of the shock. The resulting density jump ($r$) can be derived,
\begin{equation}\label{eq:r}
r\equiv \frac{n_2}{n_1} = \left\{
\begin{array}{r}
   \frac{5+\chi_1^2 \left(5-\sigma +\sqrt{\Delta}\right)}{2 \left(\chi_1^2+5\right)}, \sigma < \sigma_c, \\
  \frac{2\chi_1^2}{\chi_1^2+3}, \sigma > \sigma_c,
\end{array}
\right.
\end{equation}
where,
\begin{eqnarray}
  \chi_1^2 &=& \frac{v_1^2}{P_1/n_1},\nonumber\\
  \sigma_c &=& 1 -\frac{4}{\chi_1^2+3}-\frac{1}{\chi_1^2},\\
    \Delta &=& \frac{25}{\chi_1^4}-\frac{10 (\sigma+3)}{\chi_1^2}+(\sigma-9) (\sigma-1).\nonumber
\end{eqnarray}
The model explained in BN18 made use of the parameter $\chi_1$, which is a pseudo Mach number. The reason why the model is not adapted to the use of a proper Mach number can be understood by examining Eq.~(\ref{eq:r}), which shows that for $\sigma=0$ and $\chi_1 \gg 1$, the density jump reads $r=4$, while for $\sigma > \sigma_c$ and $\chi_1 \gg 1$, we find $r=2$. Therefore, the model presents a  transition from a 3D strong shock with $\gamma=5/3$ at $\sigma=0$, to a 1D strong shock with $\gamma=3$ at $\sigma > \sigma_c$ \citep{Bret21}. As a result, within the model of BN18, it is impossible to define a Mach number using a fixed adiabatic index since the effective adiabatic index of the model varies between 5/3 and 3.

In the rest of this work, the parameter $\chi_1$ is related to the real Mach number $\mathcal{M}$ through,
\begin{equation}\label{eq:XiM}
\chi_1^2 = \gamma \frac{v_1^2}{\gamma P_1/n_1} ~~\Rightarrow~~  \chi_1 = \sqrt{\gamma} \mathcal{M},
\end{equation}
where $\gamma$ is the adiabatic index of the plasma. 
Accordingly, the downstream  anisotropy $A_2$ reads,
\begin{equation}\label{eq:A2OK}
A_2 = \left\{
\begin{array}{r}
   \frac{1}{2}\frac{ \chi_1^2 (r-2) (r-1)+ r (5 r-3)}{\chi_1^2 (r-1)+ r}, \sigma < \sigma_c, \\
  \frac{4 \chi_1^2}{\chi_1^4 + 2 \chi_1^2-3} , \sigma > \sigma_c.
\end{array}
\right.
\end{equation}

Notably, in the strong shock limit $\chi_1=\infty$, Eqs.~(\ref{eq:r},\ref{eq:A2OK}) reduce to,
\begin{equation}\label{eq:rstrong}
r \equiv r_\infty= \left\{
\begin{array}{r}
   \frac{1}{2} \left(\sqrt{(\sigma-9) (\sigma-1)}+5-\sigma\right), \sigma < 1, \\
   2 , \sigma > 1,
\end{array}
\right.
\end{equation}
and
\begin{equation}\label{eq:A2stropng}
A_2\equiv A_{2\infty} = \left\{
\begin{array}{r}
   \frac{1}{4} \left(\sqrt{(\sigma-9) (\sigma-1)}+1-\sigma\right), \sigma < 1 \\
  0 , \sigma > 1.
\end{array}
\right.
\end{equation}
with, as expected, $r_\infty(\sigma=0)=4$ and $A_{2\infty}(\sigma=0)=1$.

\section{Simulations}\label{sec:PIC}
In order to test these predictions we perform a survey of non-relativistic 3D particle-in-cell (PIC) simulations of electron-positron (pair-plasma) shocks with TRISTAN-MP \citep{Buneman1993,Spitkovsky2005}.
The choice to model a pair-plasma was made to more directly compare with the theory developed in BN18 which was derived for an electron-positron plasma. This choice was motivated by two reasons: the temperature of the two species would have the same value downstream of the shock and the instabilities, relevant for this problem, are unchanged in the electron-positron limit \citep{Gary2009,Schlickeiser2010}.
In the ion-electron limit, the temperature of the two species is not necessarily equal \citep{Guo2017,Guo2018,Tran+20}, however, examining this problem in this framework would require a description of the ion/electron energy partition. While this is an important step in understanding the physics of non-relativistic collisionless shocks, it is beyond the scope of this work

The setup of the shock simulations are similar to those described in  \cite{Spitkovsky2008a,Sironi2011,Sironi2013}, where the shock is formed by initializing the plasma with a supersonic flow towards a reflecting wall (in the negative $x$ direction) on the left hand side of the simulation.
The plasma is reflected and forms a shock that is mediated by some counter streaming instabilities.
The plasma is continuously injected by an injector receding away from the shock which extends the simulation domain.
The boundary conditions along $y$ and $z$ are periodic.

Values in the simulations are measured in normalized code units: speeds are normalized to the speed of light, times to the inverse plasma frequency $\ompi = \sqrt{m/4\pi n e^2}$, where $e$ is the fundamental charge and $n$ is the number density of the inflowing plasma, and lengths, consistently, to the electron inertial length $d_e = \comp$.
The maximum allowed simulation length (i.e., where the receding injector stops) is $L_x = 1200 \comp$, and the transverse width and depth are $L_y = L_z = 16\comp$ for all simulations.
The simulations are performed with $2$ electron/positron pairs per cell, $4$ grid cells per $\comp$, and with the time step chosen such that $\Delta t = .45 \Delta x/c$.
A convergence simulation was run for the $M_A = M_s$ case, with double the transverse box size, twice the spatial resolution and $\times 4$ particles per cell, and no significant difference in the results found.

The inflowing electrons and positrons are initialized as a drifting Maxwellian with equal temperature, $\Delta \gamma \equiv k_BT/mc^2 = 10^{-4}$.
Because of the reflecting wall, the downstream plasma is at rest in the simulation frame.
In this frame, the upstream plasma is in-flowing at $v_{sh}/c= 0.1$,
which corresponds to a Mach number of $\mathcal{M} = v_{sh}/\sqrt{\gamma P/n} \approx 7.75$. However, the shock predictions and  theory outlined above are determined in the rest frame of the shock.
The two frames are ultimately related by the density jump $r$, i.e., $v_1 = v_{sh}/(1 - 1/r)$, as discussed in many previous simulations of shocks formed by reflecting walls \citep{Caprioli2014,Haggerty2019,haggerty+2020}.

The simulations are initialized with a uniform magnetic field anti-parallel to the upstream flow (i.e., $+x$).
The initial magnetic field strength varies between simulations and is characterized by the shock energy density in the downstream frame $\sigp = B_1^2/4\pi n_1 m v_{sh}^2$ and is related to the value described by Eq.~\ref{eq:sigma} by the density jump, $\sigma = \sigp(1 - 1/r)^2$.
The survey consists of 9 simulations performed over a range of $\sigp$ values from $10^{-3}$ to $25$, allowing for the consideration of both super-Alfv\'enic and sub-Alfv\'enic shocks.
The simulations are evolved for 1000's of $\ompi$ but are stopped before particles can escape the simulation.

\section{Hydrodynamics}\label{sec:hydro}
Various hydrodynamic shock quantities can be determined from the survey of simulations, and in the following section they are presented and compared with the theory of BN18. 
\subsection{Density Jump}
\begin{figure}
 \includegraphics[width=\columnwidth]{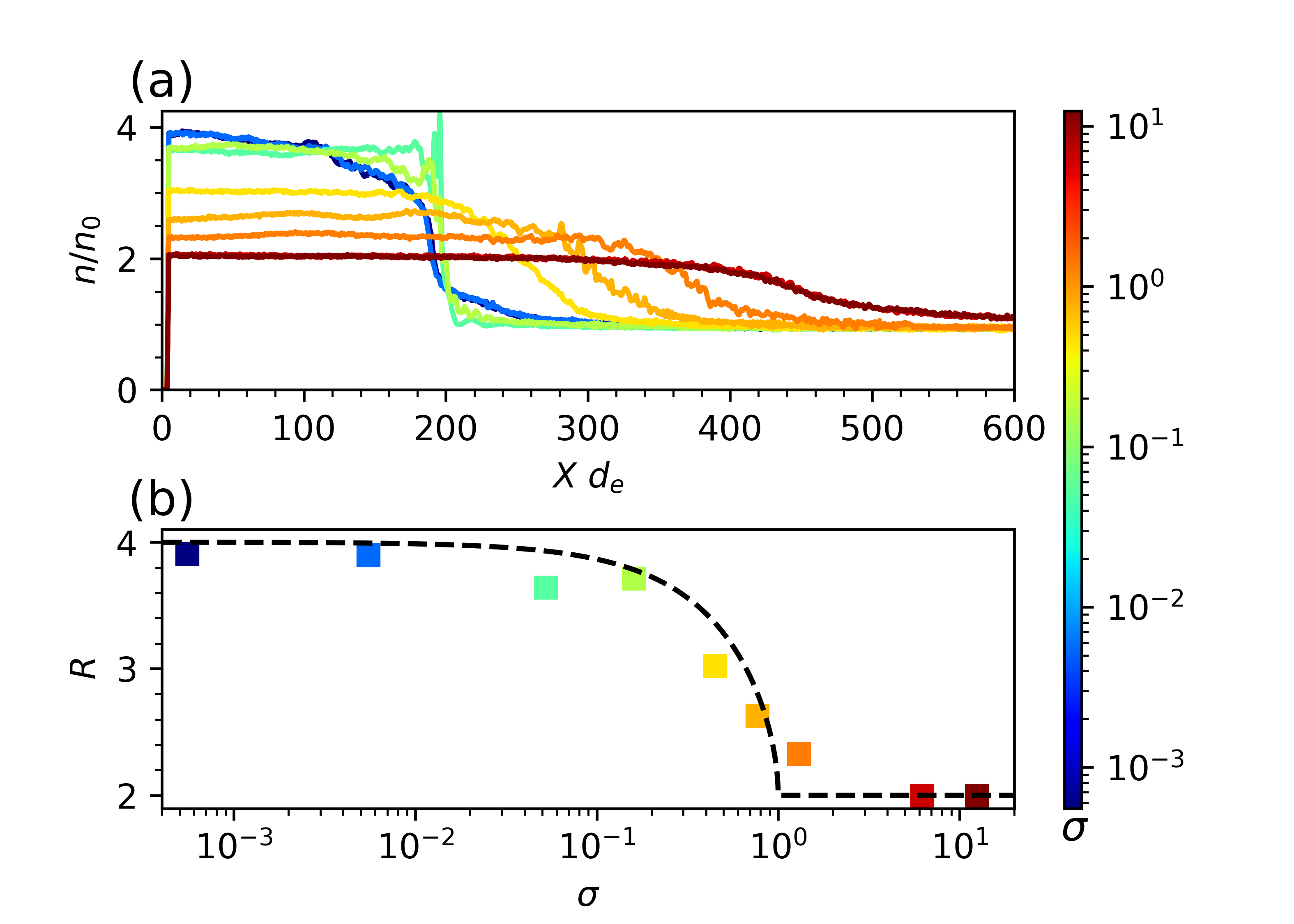}
 \caption{Panel~(a): Shock density profile integrated in the transverse directions, for  $\chi_1=10$ and various $\sigma$, for this and all following figures, the data points are color-coded by value of $\sigma$ and correspond to the values in the color bar. Panel~(b): The average downstream density normalized to the upstream density for various $\sigma$. Data are plotted at time  $6000 \ompi$ and the dashed line pertains to Eq.~(\ref{eq:rstrong}).}
 \label{fig:density}
\end{figure}

Fig.~\ref{fig:density} shows the shock density profile integrated in the transverse directions, for $\sigma$ up to 25. Data are plotted at time  $6000 \ompi$ to insure that the shock is fully formed.

The jump obtained for values of $\sigma \ll 1$ is 4 consistent with the MHD prediction for strong shocks.
Although the critical magnetization is close to unity ($\sigma_c \sim 1$), a clear departure from the standard MHD prediction 
is observed for $\sigma$ small as $0.1$.
In this regime, where $0.1 \lesssim \sigma \lesssim 1$, the Mach number based on the Alfv\'en speed would range between $1$ and $10$, and so the decreasing compression ratios seems consistent with the decreasing Mach number.

The lower plot in figure \ref{fig:density}  displays the density jump for a range of $\sigma$ values, contrasted against Eq.~(\ref{eq:rstrong}) represented by the black dashed line. 
Simulations show the downstream  density enhancement decreasing from 4 to 2, consistent with prediction presented above.
The theoretical predictions attribute the departure from the MHD prediction to the stabilization of downstream temperature anisotropy, which can be directly examined form the summations and is discussed in Sec.~\ref{sec:aniso}

\begin{figure}
 \includegraphics[width=\columnwidth]{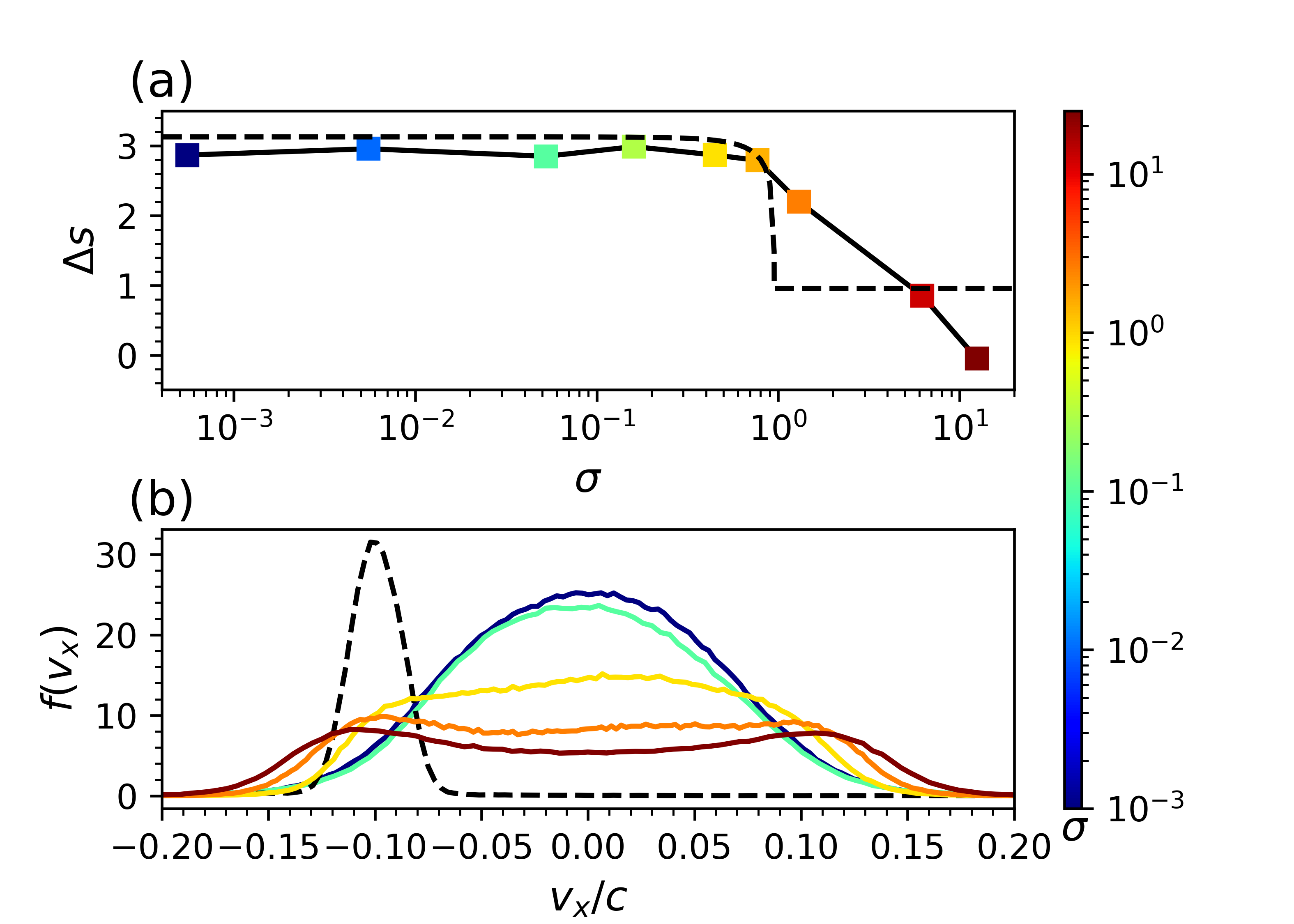}
 \caption{Panel (a): The change in the kinetic entropy, as defined by Eq.~\ref{eq:entropy}, across the shock in the simulations with different values of $\sigma$ (color coded). Panel (b): The downstream parallel  distribution function ($\int f dv_ydv_z$) from simulations with different values of $\sigma$. The upstream distribution shown by the black dashed line at time $6000 \ompi$. Simulations with smaller values of $\sigma$ show larger entropy increases with more normal downstream distributions, while simulations with larger $\sigma$ values have a smaller entropy increase, with distributions that are double peeked.}
 \label{fig:entropy}
\end{figure}

\subsection{Entropy}
\begin{figure}
 \includegraphics[width=\columnwidth]{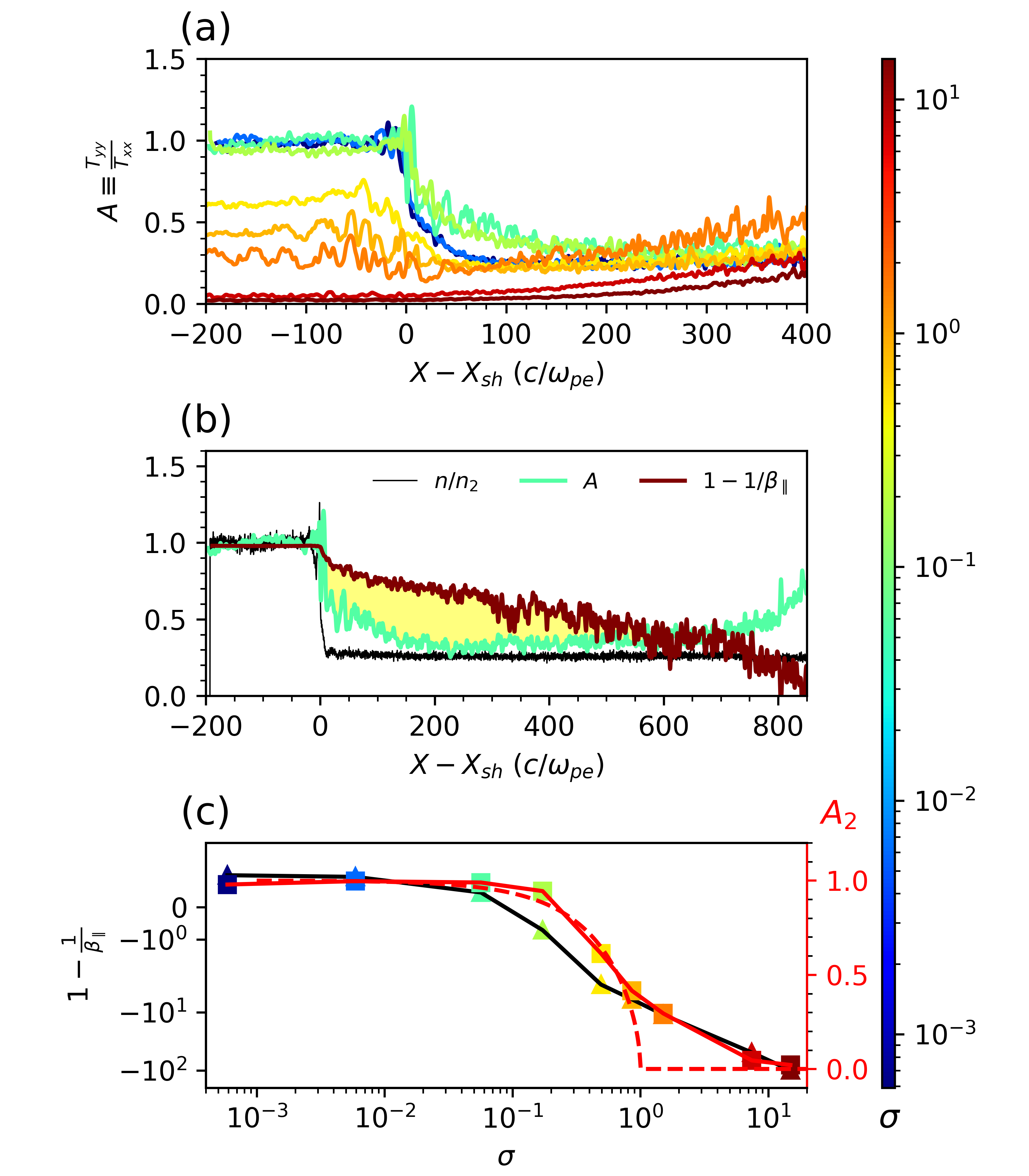}
 \caption{Profiles of anisotropic parameters for strong shocks ($\chi_1 = 10$), taken at $6000 \ompi$, with varying values of $\sigma$ (color-coded).
Panel~(a): Ratios of the perpendicular to parallel electron temperature ($A=T_{yy}/T_{xx}$), averaged over the transverse directions and shown relative to the shock front.
Panel~(b): Profiles of both sides of Eq.~\ref{eq:firehose} (left side, $A$, green line; right side, firehose, maroon) superimposed on the density normalized to the downstream value (black line) for $\sigp=0.1$ ($\sigma \approx 0.056$). The system is firehose unstable in regions where $1 - 1/\beta_\parallel > A_2$, corresponding to the yellow shaded region in the upstream, in contrast to the marginal stability in the downstream.
Panel~(c): The averaged downstream temperature anisotropy (right axis, red line with squares) and firehose parameter (left axis, black line with triangles) shown as a function of $\sigma$. The red dashed line correspond to prediction for $A_2$ in the strong shock limit, as defined in Eq.~\ref{eq:A2OK}. 
}
 \label{fig:firehose_strong}
\end{figure}

For simulations with relatively small values of $\sigma \ll \sigma_c$, it is clear that a shock in the typical sense is forming, i.e., the upstream and downstream regions are separated by a sharp discontinuity, and the downstream plasma is compressed consistent with the MHD jump conditions; however, for simulations with $\sigma \gtrsim \sigma_c$ it is less clear if a {\it bona fide} shock has developed, or 
we are witnessing interpenetrating fluids. 
Ultimately, shocks must increase entropy, and as such the issue can be addressed by the entropy difference between the upstream and the downstream. 
While there remain unsettled issues in the interpretation of entropy and reversibility in a collisionless system, or even effective collisions introduced by poor counting statistics of the PIC model \citep[e.g.][and references therein]{Montgomery+70,Shalaby+17,Kadanoff17,Yang+17,Liang+19,Du+20}, we limit considerations to the definition of kinetic entropy density per particle:
\begin{equation}\label{eq:entropy}
    s = \frac{S}{n} = -\frac{\int f\ln f d^3v}{\int f d^3v}
\end{equation}
where $S$ is the kinetic entropy density, $f$ is the distribution function.
This calculation was explicitly performed in Sec.~4 of BN18 and considered again in \cite{Bret21} where the proof is made that for large values of $\chi_1$, the entropy jump 
for both the large and small $\sigma$ regimes is definitely positive. 
Since the marginal firehose stability is observed in the present simulations at small $\sigma$ (see Fig. \ref{fig:firehose_strong}), while the perpendicular temperature conservation is also observed far larger $\sigma$'s (see Sec~\ref{sec:aniso}), the calculations of BN18 and \cite{Bret21}  should apply, ensuring an entropy increase at the transition.
The measured entropy jump ($\Delta s$) for all of the simulations as a function of $\sigma$ is shown in panel (a) of Fig.~\ref{fig:entropy}, with the black dashed line showing the theoretical prediction.
The value is calculated by taking the difference of integrated discretized distribution functions averaged over a region of $100\ d_e$ in the downstream and upstream. 
There is good quantitative agreement for simulations with $\sigma < \sigma_c$, 
and as  $\sigma$ is increased and the downstream remains firehose stable across the shock, there is qualitative agreement with theory, i.e., the difference in entropy reduces.
It is worth noting that the entropy increase for the  $\sigma \approx 6$ simulation agrees well with theory, however, in the simulation with an even larger magnetization, $\sigma \approx 14$, the entropy goes to nearly zero ($\Delta s \approx 0$).
This difference can be understood by considering the parallel downstream distribution functions for these simulations, shown in panel (b) of Fig.~\ref{fig:entropy}.
In the entropy derivation of BN18, the distribution was assumed to have the form of a bi-maxwellian, while it is clear from the dark red line in the lower panel, that the distribution has two clear peaks associated with counter-streaming beams.
However, it is likely that ---if the simulation were run for longer times in a larger box--- the double-peaked distribution function would tend towards the bi-maxwellian state, as the distribution function has already gaussianized relative to the initial distribution.
From this it seems reasonable that the entropy will increase across the discontinuity, thus constituting a collisionless shock even in large-$\sigma$ systems.

\subsection{Firehose stability}
The density jump measured in the simulations is in good agreement with the theoretical predictions from BN18 and to further verify the consistency between simulations an BN18, we also test the assumptions that form the basis of the model.
The main ingredient of the BN18 model is that for weak magnetic field strengths, the plasma turns firehose unstable at the shock front crossing. As a result, it migrates to the stability threshold so that the downstream is marginally firehose stable. For strong fields, we expect the downstream to never becomes unstable, the condition for which is \citep{Gary2009,Schlickeiser2010},
\begin{equation}\label{eq:firehose}
 A_2 > 1 - \frac{1}{\beta_{\parallel 2}},
\end{equation}
 with,
\begin{equation}\label{eq:beta}
 \beta_{\parallel 2} = \frac{n_2k_BT_{\parallel 2}}{B_0^2/8\pi} = \frac{n_2k_BT_{x 2}}{B_0^2/8\pi} = \frac{4}{\chi_2^2\sigma}.
\end{equation}

Fig.~\ref{fig:firehose_strong}(a) shows the anisotropy parameter $A(x)=T_y/T_x$ with the x-axis re-centered on the shock front for various values of  $\sigma$ (color coded).  
The anisotropy far upstream for all simulations, namely $x \gtrsim 700 \comp$, is 1 but it decreases in the shock foot\footnote{In order to correctly render the shock front and the downstream, we do not show this far upstream where $A$ reaches unity}. 
In contrast with what was assumed in  BN18, the structure of the shock is not a discontinuity, which will be discussed further in Sec.~\ref{sec:width}.
The anisotropy parameter downstream behaves as expected; at low $\sigma$, the field cannot sustain a stable anisotropy and the downstream is nearly isotropic and $A_2 \sim 1$. As the field strength increases and $\sigma > \sigma_c$, $A_2$ progressively goes to 0.
This is further demonstrated in Fig.~\ref{fig:firehose_strong}(c), in which the red line (corresponding to the right y-axis) shows the average downstream value of $A_2$ as a function of $\sigma$.
In Fig.~\ref{fig:firehose_strong}(c) the red dashed line shows the prediction for $A_2$ as a function of $\sigma$ (Eq.~\ref{eq:A2stropng}), and the two are found to be in good agreement.

Another ingredient of BN18 was that for weak fields, the plasma turns firehose unstable when crossing the shock front before it migrates to marginal stability. In this respect, Fig.~\ref{fig:firehose_strong}(b) shows both sides of Eq.~\ref{eq:firehose} superimposed on the shock density profile for $\sigma=0.056$ ($\sigp = 0.1$). While the downstream shows marginal stability with $ A_2 \sim 1 - 1/\beta_{\parallel 2}$, we find $ A_2 < 1 - 1/\beta_{\parallel 2}$ for $100 \comp < x < 250 \comp$. As expected, the plasma turns firehose unstable as it passes to the downstream, before it settles to marginal stability.

Finally, we can determine the downstream firehose condition for the different $\sigma$ simulations; Fig.~\ref{fig:firehose_strong}(c) shows the averaged downstream values from the two sides of Eq.~\ref{eq:firehose} (with $A_2$ denoted by the red line with squares measured by the right axis and the black line with triangles measured by the left axis). For small values of $\sigma $ Both terms are close to 1. As $\sigma$ increases to $\sigma \gtrsim 0.1$, the LHS of Eq.~\ref{eq:firehose} drops and even becomes negative implying that the downstream is firehose stable, while simultaneously the temperature ratio ($A_2)$, also decreases from 1 to 0 showing that a stable field aligned temperature anistropy has developed.

\subsection{Structure of the Downstream Temperature Anisotropy}\label{sec:aniso}
\begin{figure}
 \includegraphics[width=\columnwidth]{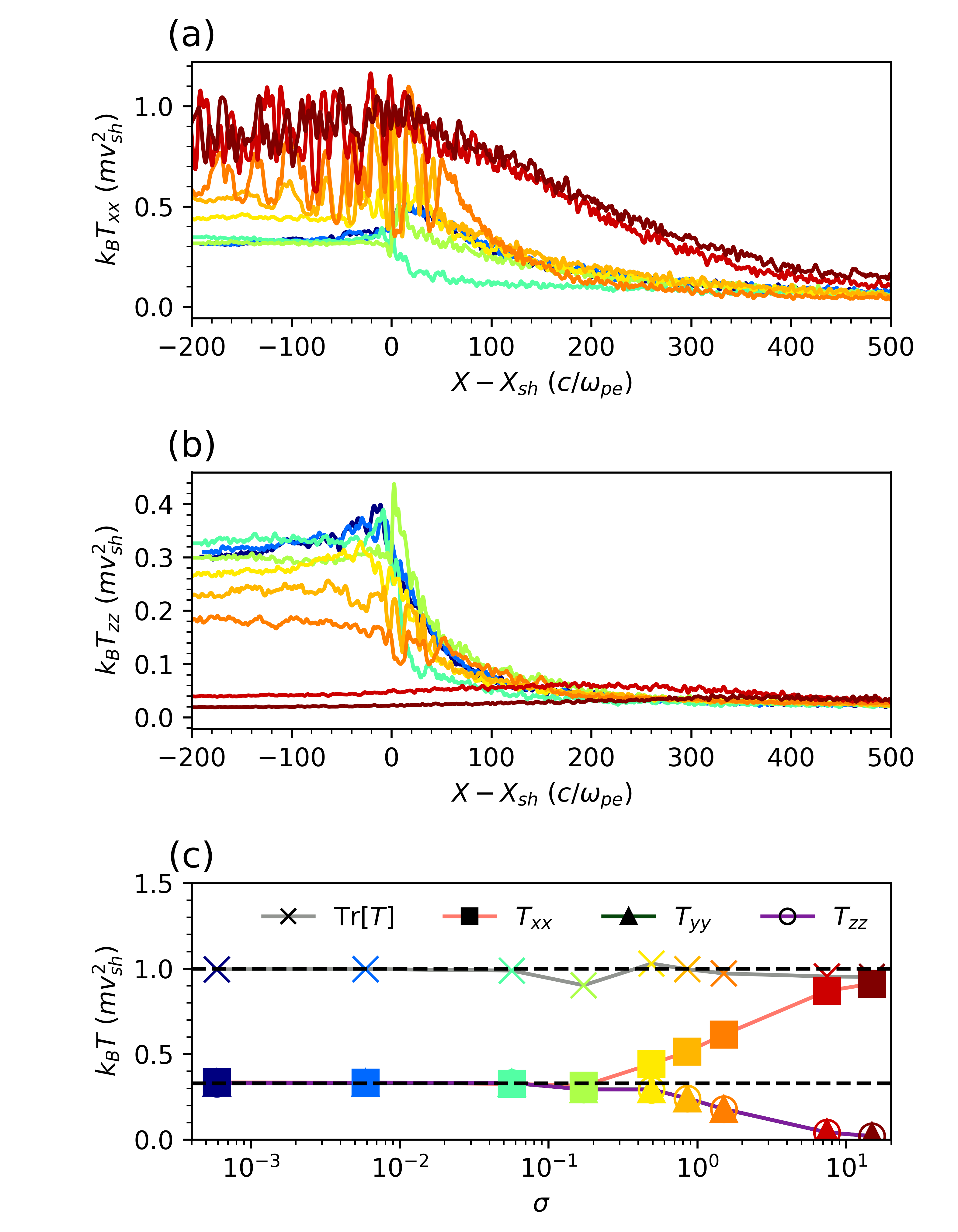}
 \caption{Panel~(a) and (b): Profiles of the parallel ($T_{xx}$, (a)) and perpendicular ($T_{zz}, (b)$) temperatures, averaged along the transverse directions, showing the development of a strong temperature anisotropy for increasing values of $\sigma$, color-coded.
 Panel~(c): Components of the average downstream temperatures, in units normalized to $mv^2_{1} = mv_A^2/\sigma$, as a function of $\sigma$, with the two black dashed lines corresponding to $1$ and $1/3$ respectively, with $T_{xx}$ as pink, $T_{yy}$ as dark green, $T_{zz}$ as purple and the trace of the temperature tensor, $\mathrm{Tr}[T] = \sum_i T_i$ as grey.}
 \label{fig:temp}
\end{figure}

The predictions laid out in BN18 ultimately derived from the temperature anisotropy generated downstream of the shock.
The anisotropy is predicted to grow for strong magnetic fields, as the inflowing plasma is expected to remain magnetized, such that the perpendicular temperature should not change appreciably across the shock.
This assumption is examined in Fig.~\ref{fig:temp}, which shows the parallel ($T_{xx}$, panel a) and out-of-plane ($T_{zz}$, panel b) temperature profiles for the different shock simulations.
Note that the temperatures in this figure are normalized to twice the kinetic energy of the inflowing plasma, rather than to the rest mass energy, and the x-axis is centered around the shock.
For the simulations with smaller values of $\sigma$ ($\lesssim 0.3$, blue to green) both the parallel and perpendicular temperatures increase to roughly equal values downstream, resulting in an isotropic downstream temperature, consistent with Fig.~\ref{fig:firehose_strong}(a). For increasing magnetization ($\sigma \gtrsim 0.3$, yellow to red), The parallel temperature increases dramatically, while the perpendicular temperature become nearly constant across the shock, verifying the assumption of BN18.

Further details about the downstream anisotropy and be found from Fig.~\ref{fig:temp}(c), which analyses how the thermal energy is split among the degrees of freedom.
In the plot, the average downstream temperatures are shown as a function of $\sigma$, with the parallel temperature represented by the pink line with squares, perpendicular with a green line with triangles (y)/purple line with circles (z) and the average with a gray line with `x's (denoted by $\mathrm{Tr}[T]=\sum T_i$.)
\footnote{The choice of normalization makes it clear that the total downstream temperature is equal to the second moment of two counter streaming beam populations moving at $\pm v_{sh}$.
This fact is likely a consequence of the simulation design, as the downstream population must be at rest in simulation frame, and so all of the energy in this frame must be thermal.
It seems natural that the downstream temperature must be equal to the composite energy of the counter streaming cold beams.}
For $\sigma < 0.1$, the different components of the temperature are equal.
As the only stable configuration has $A_2\sim 1$, the downstream is isotropic with its thermal energy evenly distributed along the 3 degrees of freedom.
As a consequence, each one carries 1/3 of the total, and all together they sum up to $mv^2_{sh}$  which is the total amount of energy initially available in the 2 cold counter streaming plasmas.
For $\sigma > 0.1$, the downstream becomes increasingly anisotropic with increasing $\sigma$ according to the mechanism described in BN18.
The parallel temperature $T_x$ rises while $T_y$ and $T_z$ decrease  in such a way that $\mathrm{Tr}[T]$ is kept constant and equal to $mv^2_{sh}$ (momentum conservation).

\section{Shock Front Width}\label{sec:width}
\begin{figure}
 \includegraphics[width=\columnwidth]{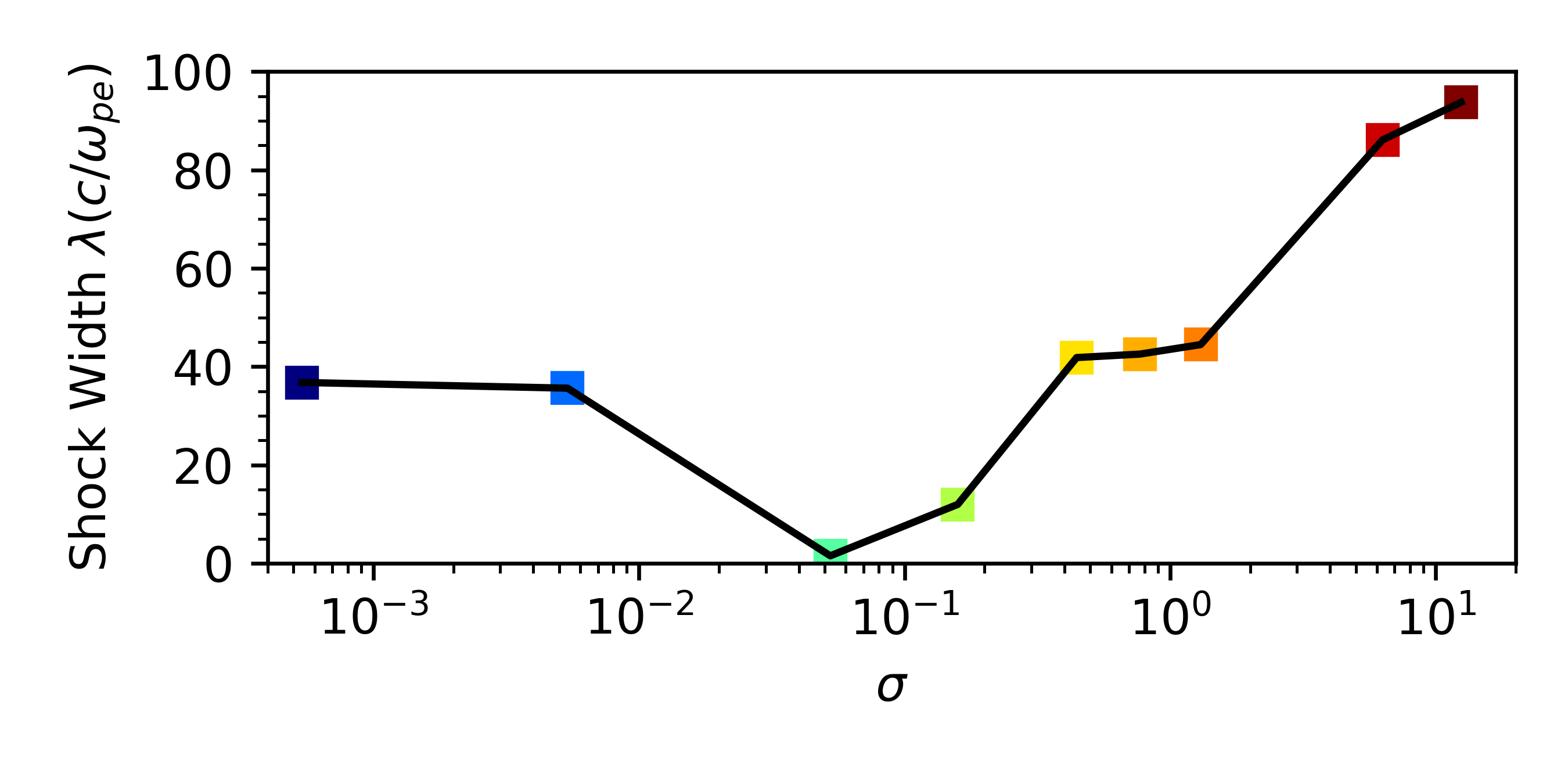}
 \caption{The width of the shock as a function of $\sigma$, determined from the best fit of a ${\rm tanh}$ function shown in Eq.~(\ref{eq:tanh}), color-coded by $\sigma$.}
 \label{fig:widths}
\end{figure}

 From the density profiles shown in Fig.~\ref{fig:density}(a), it is clear that the width of the shock has some dependence on $\sigma$.
 To further examine this effect, we determined the width of the shock fronts for each simulation at $6000\, \ompi$. This was achieved by fitting the shifted, average density profiles with a ${\rm tanh}$ function of the form,
 \begin{equation}\label{eq:tanh}
     \frac{n(x)}{n_1} = \frac{1+r}{2}+\frac{r-1}{2}\tanh(-x/\lambda).
 \end{equation}
 The value of $\lambda$ for each simulation is determined with a non-linear least squares fit and the results are shown in Fig.~\ref{fig:widths}.

 Although the width of the shock varies non-monotonically with $\sigma$, two distinct patterns appear to emerge when considering simulations with $\sigma \ll 0.1$ or $\sigma \gg 0.1$. In the small $\sigma$ case, the shock width seems to reach a constant size for decreasing $\sigma$.
This is consistent with the shock formation being mediated by the two stream instability for the present parameters \citep{BretPRL2008}, which has a growth rate independent of $\sigma$.

For $\sigma \gg 0.1$, the shock width increases with $\sigma$.
This increase is potentially due to the increasing number of pitch angle scattering required to return particles traveling up-streaming back towards the downstream.
Each deflection is expected to alter a particle's pitch angle by $\Delta mu \sim \delta B/B_0 \sim 1/\sqrt{\sigma}$.
For a larger value of $\sigma$, more deflections will be required to send particles back potentially leading to a larger shock width.
This process is shown in an animation of self-consistent particle tracing for the $\sigma^\prime = 2$ simulation (orange line in this papers figures) in the supplementary material. 

\section{Conclusion}\label{sec:conc}
In this work, using self-consistent kinetic PIC simulations, we verify the departure of parallel collisionless shocks from fluid MHD predictions; by varying the magnetization on collisionless pair plasma shocks we show that a parallel propagating shock wave is suppressed for sufficiently strong magnetic field strengths, consistent with kinetic theoretical predictions of BN18.
For a fixed (parallel) shock inclination and sonic Mach number ($v_1/v_{th} = 10$), we test the predictions of BN18 of how the shock properties depend on the strength of the upstream magnetic field relative to the upstream shock speed, a quantity parametrized by $\sigma \equiv B_0^2/4\pi n_1 m v_1^2$.
The simulations are found to be in good agreement with the theoretical predictions, namely that as $\sigma$ is increased to $\gtrsim 1$, the shocked plasma remains well magnetized, suppressing the firehose instability and leading to a large field aligned temperature anisotropy downstream. The entropy jump between the downstream and upstream is considered and found to be approximate agreement with predictions.

Additionally we examine how the kinetic properties of the shock are modified by the magnetization. The effective shock width is determined for a fixed time for each of the simulations, and the width is found to be a minimum for an upstream plasma beta of unity ($\beta \sim 1$). For decreasing and increasing values of $\sigma$ the shock width broadens, but saturating for small values of $\sigma$ as the instability mediating the shock reformation becomes two stream like and insensitive to the magnetic field strength.

In the theory and simulations of this work, only a pair plasma was considered; this choice simplified the problem by ensuring that the downstream temperatures would be the same for both species. In shocks comprised of electron-proton plasmas, it is likely that the temperature of each species will be different \citep{Tran+20}. This added complexity should be further explored to insure the applicability of this work to space and astrophysical shocks and will be considered in future investigations.

These results contribute to the building consensus collisionless shocks fundamentally require a kinetic description to accurately capture both the thermal and non-thermal plasma physics. 

\section*{Data Availability}
The datasets used in this manuscript were derived using the open source particle-in-cell software {\it Tristan-MP v2} (\url{https://ntoles.github.io/tristan-wiki/}).
The the average output for each simulation requires $\sim 14$ GB of space ($\gtrsim 200$ GB for all of the simulations and convergence studies) and as such either the data underlying this article or the simulation initialization files will be shared on reasonable request to the corresponding author.

\section*{Acknowledgements}
We thank the anonymous referee for thorough comments which helped improve the manuscript.
This work has been achieved under projects ENE2016-75703-R from the Spanish Ministerio de Econom\'ia y
Competitividad and SBPLY/17/180501/000264 from the
Junta de Comunidades de Castilla-La Mancha.
As well as the FDSS NSF AGS-1936393 grant to the Institute for Astronomy of the University of Hawaii.
Simulations were performed on computational resources provided by the University of Chicago Research Computing Center and XSEDE TACC (TG-AST180008).

\bibliographystyle{mnras}
\bibliography{BibBret} 

\bsp	
\label{lastpage}
\end{document}